\documentclass[twocolumn,showpacs,preprintnumbers,amsmath,amssymb]{revtex4}

\usepackage{graphicx}
\usepackage{dcolumn}
\usepackage{bm}

\begin{document}


\title{Fractality feature in oil price fluctuations}

\author{M. Momeni}
 \email{momeni@shahroodut.ac.ir}
\affiliation{%
Faculty of Physics, Shahrood University of Technology, Shahrood,
Iran
}%

\author{I. Kourakis}
 \affiliation{Queen's University Belfast,
Center for Plasma Physics, BT7 1 NN Northern Ireland, UK}

\author{K. Talebi}

\affiliation{Faculty of mining  engineering , Shahrood University of
Technology, Shahrood, Iran}

\date{\today}

\begin{abstract}
The  scaling properties of  oil price fluctuations are described
as a non-stationary stochastic process realized by a time series
of finite length.  An original model is used to extract the
scaling exponent of the fluctuation functions within a
non-stationary process formulation. It is shown that, when returns
are measured over intervals less than 10 days, the Probability
Density Functions (PDFs) exhibit self-similarity and monoscaling,
in contrast to the multifractal behavior of  the PDFs at
macro-scales (typically larger than one month). We find that the
time evolution of the distributions are well fitted by a L\'evy
distribution law at micro-scales. The relevance of a L\'evy
distribution is made plausible by a simple model of nonlinear
transfer.
\end{abstract}


\maketitle

\section{introduction}

Many natural or man-made phenomena, such as turbulence flows,
fluctuations in finance (stock market), seismic recording,
internet traffic, climate change, etc., are characterized by
randomness or stochasticity \cite{Fris, bisk, Kolm1,noroz, mant,
mant1, bouch, abra, farah,ken,keil, vere, voig, bufe,
saic,kosc,barb,frae}. Analysis of non-stationary stochastic
processes referring to quantities which fluctuate widely and are
uncertain has been a problem of fundamental interest for a long
time.  Over the past two decades, oil price has increased very
sharply, rising from \$25 per barrel in January 1986 to a peak of
close to \$ 122 per barrel in the last week of July  2008. The
effects of oil price fluctuations on the world economy are
undeniable and particularly evident from the international
reports. Oil price data as a time series exhibit complex patterns
and seemingly appear to be a chaotic system. Indeed, the behavior
of oil price fluctuations can be efficiently modelled by the Ising
model which was proposed for stock-price fluctuations \cite{pler}
and by the the cascade model developed based on fractal concepts
which was used for hydrodynamics and magnetohydrodynamic
turbulence \cite{ghas,muzy,mom}. Here, we employ the latter
technique
 to characterize the statistical properties of the oil-price time series,
which sensitively distinguish between self-similarity and
multi-fractality  in a time series.  The model is based on
two-point increments of  oil price, yielding a comprehensive and
scale-dependent characterization of the statistical properties of
the system  via an associated Probability Density Function(PDF).
It is necessary to stress that the data series
 is represented by a finite number of records which  do not constitute
 a stationary process. The effect of non-stationarity on the
 detrended fluctuation analysis has been investigated in Ref. \cite{hu}. To
 eliminate the effect of sinusoidal trend, we apply the Fourier
 Detrended Fluctuation Analysis (F-DFA). After the elimination of
 the trend we use the Multifractal Detrended Fluctuation Analysis
 (MF-DFA) to analysis the data set. The MF-DFA methods are the
 modified version of detrended fluctuation analysis (DFA) to detect
 multifractal properties of time series. The detrended fluctuation
 analysis (DFA) method introduced by Peng et al. \cite{peng} has became a
 widely-used technique for the determination of (multi)fractal
 scaling properties and the detection of long-range correlations in
 noisy, non-stationary time series \cite{hu, peng}. It has successfully been
 applied to diverse fields such as DNA sequences \cite{buld,buld1}, heart rate
 dynamics \cite{ken, ken1}, neuron spiking \cite{bles}, human gait \cite{haus}, long-time weather
 records \cite{kosc, ivan}, cloud structure \cite{ivan1}, geology \cite{mala}, ethnology \cite{alad}, economical
 time series \cite{mant,liu}, solid state physics \cite{kant}, sunspot time series \cite{sadegh}, and
 cosmic microwave background radiation \cite{keil, vere}.

 We propose a method for generating a stationary process analysis out of
 a non-stationary process. The fact is that  stationary stochastic systems often show scaling in a
statistical sense, coincident with non-Gaussian  leptokurtic
(heavy- tailed) statistics. Importantly, identification of the
associated scaling exponents implies the ability to interpret and
estimate the behavior of the fluctuations as well as the detection
of long-range correlations. A self-similar Brownian walk with
Gaussian PDFs, which has scaling exponent $1/2$,  is a good
example of the process where  shows uncorrelation on all temporal
scales. We try to determine the
 scaling  properties of the PDFs  that are
 leptokurtic at micro-scales. The scaling exponents can be determined through the
 scaling behavior of the moments, usually characterized by computing
 structure functions.   It is said that the fluctuations are  self-similar (monofractal) if scaling exponents of the  moments exhibit a linear
  power-law dependence.
 In contrast a nonlinear dependence  infers to  multifractal scaling, which is caused by intermittent small-scale structures of oil price fluctuations.
 A   similar feature has been  found in  physical systems for example, in velocity and magnetic fields of the  solar
 wind \cite{kian,kian1,gold} as well as in magnetohydrodynamic turbulence  studied via  direct numerical
 simulations \cite{mom}.
  Finally,  distributed price changes are characterized by a
  stable L\'evy distribution in the central part of the distribution.

 The paper is structured as follows. In Section II we describe our
 data set. The
 MF-DFA method is briefly presented in Section  III
 and shown  that the scaling exponent determined via
 the MF-DFA method are identical to those obtained by the standard
 multifractal formalism based on PDF
 analysis.  In Section IV we employ a recently  developed technique \cite{kian,kian1,mom} that sensitively
 distinguishes between self-similarity and multifractality in times series.
  By analyzing the temporal evolution of  price
 dynamics, we demonstrate the strongly the non-Gaussian  behavior of
 the returns of  oil price and scale-dependent  behavior of the PDFs.
 Also we explain the Hurst exponents analysis.
 The micro-scale PDFs resemble leptokurtic L\'evy distribution which
 will be discussed in Section V. Finally, in Section VI we will
 summarize all results discussed throughout this paper.

\section{The data}

\begin{figure}
\includegraphics[width=0.4\textwidth]{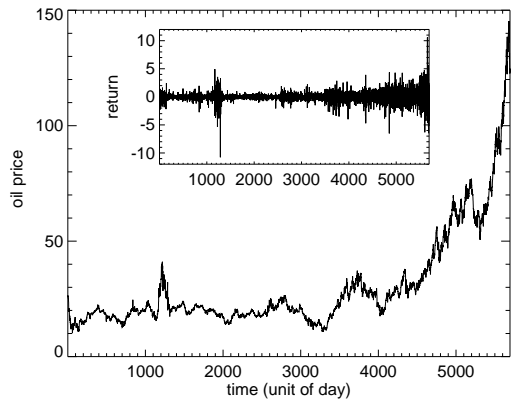}
\caption{\label{p_re}Semi-log plot of oil price series over the
period 1986-2008. Inset:  daily return of the oil price index.}
\end{figure}

Over a twelve-year period, on average, the price of  oil has
increased from \$25 per barrel in January 1986 to a peak of close
to \$ 122 per barrel in the last week of July  2008 . Oil price as
recorded in international markets \cite{sit} offer us an almost
unique possibility to gain information on the stochastic dynamics
state in a very large scale range, say 1 day up to 200 days. Our
database consists of about  5695 daily price values which seem to
provide a set of data points which will be sufficient to obtain
the scaling properties the system. Fig.\ref{p_re} presents the
daily fluctuations in  oil price $p(t)$ in the period 1986-2008.
It is evident from the figure that the fluctuations do not
constitute a stationary process, for instance one can show that
the variance of the signal in some window does not remain stable
upon increasing the window size. Let us introduce the increments
(or returns)  $\delta p(t,\tau)$ defined by,  $\delta
p(t,\tau)=p(t+\tau)-p(t)$. The resulting series for $\delta
p(\tau)$ is shown in the inset graph of Fig.\ref{p_re}.
It is straightforward to show, by measuring the average and the
variance of $\delta p(t,\tau)$ in a moving window, that  $\delta
p(t,\tau)$ is stationary. Upon initiating the  analysis of the
distribution of oil price returns, the mean, standard deviation,
skewness, and kurtosis of the return series are calculated (see
Table \ref{table}). It is easy to show that the skewness of a
Gaussian function is zero so that the negative value of skewness,
$\lambda=-0.606749$, is a hallmark of departure of the PDFs from
the Gaussian distribution (such as a leptokurtic distribution),
which confirms the existence of intermittency in the fluctuations.
On the other hand the large value of kurtosis, $\kappa=9.55225$,
with respect to Gaussian kurtosis $(\kappa=3)$, show that the
tails of the return distribution are fatter than the Gaussian
ones.
\begin{table} \caption{\label{table}Mean, standard deviation, skewness, and kurtosis of the oil price returns.  }
\begin{ruledtabular}
\begin{tabular}{cccc}
 Mean & Standard Deviation & Skewness & Kurtosis \\
\hline
 0.0124274 &  0.729173
 &   -0.606749 &   9.55225
\\
\end{tabular}
\end{ruledtabular}
\end{table}

\section{The MF-DFA analysis}\label{MF}

 The MF-DFA method is a
modified version of detrended fluctuation analysis to detect
multifractal properties of a time series. Omitting unnecessary details, a brief summary of the method for
calculating MF-DFA  based on fractal  concepts can be formulated in five
steps. We take the price series $p_{k}$ with the size of $N$ and
follow the steps as follow:

$\bullet$ \textit{step 1:} Determine
the " profile"
\begin{equation}
Y(i)=\sum_{k=1}^{i}[p_{k}-\langle p \rangle], \hspace{1cm}
i=1,....,N\, \label{step1}
\end{equation}
where $\langle p \rangle$ is the mean  of the series. Subtraction of
the mean $\langle p \rangle$ is not compulsory, since it would be
eliminated by the detrending later in the third step.

$\bullet$
\textit{step 2:} Divide the profile $Y(i)$ into $N_{s}\equiv
int(N/s)$ nonoverlapping segments of equal lengths $s$. Since the
length $N$ of the series is often not a multiple of the considered
time scale $s$, a short part at the end of the profile may remain.
In order not to disregard this part of the series , the same
procedure is repeated starting from the opposite end.

$\bullet$
\textit{step 3:} Calculate the local trend for each of the $2N_{s}$
segments by a least-square fit of the series. Then determine the
variance
\begin{equation}
F^{2}(s,\nu)\equiv
\frac{1}{s}\sum_{i=1}^{s}\{Y[(\nu-1)s+i]-y_{\nu}(i)\}^{2},\
\label{step3}
\end{equation}
for each segment $\nu$, $\nu=1,...,N_{s}$ and
\begin{equation}
F^{2}(s,\nu)\equiv
\frac{1}{s}\sum_{i=1}^{s}\{Y[N-(\nu-N_{s})s+i]-y_{\nu}(i)\}^{2},\
\label{step4}
\end{equation}
for $\nu=N_{s}+1,...,2N_{s}$. Here we use  the linear fitting
polynomial $y_{\nu}(i)$ in segment $\nu$.

$\bullet$
\textit{step 4:} Average over all segments to obtain the $q$-th
order fluctuation function, defined as:
\begin{equation}
F_{q}(s)\equiv
\biggl\{\frac{1}{2N_{s}}\sum_{\nu=1}^{2N_{s}}[F^{2}(s,\nu)]^{q/2}\biggr\}^{1/q}
,\ \label{step4}
\end{equation}
where, in general, the index variable $q$ can take any real value
except zero. We repeat steps 2, 3 and 4 on several timescales $s$. It
is apparent that $F_{q}(s)$ will increase with increasing $s$.

$\bullet$
\textit{step 5:} Determine the scaling behavior of the fluctuation
functions by analyzing  log-log plots of $F_{q}(s)$ versus $s$.

\begin{figure}
\includegraphics[width=0.5\textwidth]{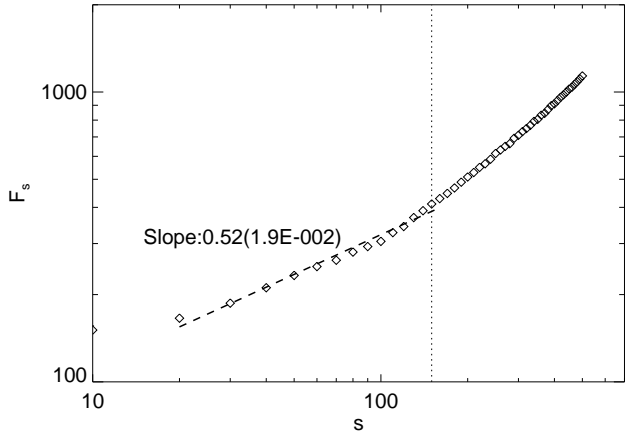}
\caption{\label{f_s} The fluctuation function $F_{2}(s)$ as a function
of box size $s$ for the returns of the oil index.}
\end{figure}

A power law relation between $F_{q}(s)$ and $s$ indicates the
presence of scaling: $F_{q}(s) \sim s^{\alpha(q)}$. In general, the
exponent $\alpha(q)$  may depend on $q$. For stationary time series
such as fGn (fractional Gaussian noise), $Y (i)$ in Eq.
(\ref{step1}), will be a fBm (fractional Brownian motion) signal,
so, $0<\alpha(q=2)<1.0$ \cite{sadegh1}.
 As a result,   the fluctuation function $F_{2}(s)$  shows a scaling behavior, $\alpha(2)$, which is
identical to the well-known Hurst exponent $H$. The Hurst exponent
is called the scaling exponent or correlation exponent, and
represents the correlation properties of the signal. If $H=0.5$,
there is no correlation and the signal is an uncorrelated signal, if
$H<0.5$, the signal is anticorrelated, if $H>0.5$, there is positive
correlation in the signal. We obtain the following estimate for the
Hurst exponent, $H=0.52\pm0.02$ as we can see in Fig. (\ref{f_s}).
Since $H>0.5$ it is concluded that  oil price returns show
persistence, i.e, a certain correlation among consecutive
increments. For
$s\sim 150$ the empirical data deviate from the initial scaling
behavior, as we can see in Fig. (\ref{f_s}). This indicates that oil
price tends to loose its memory after a period of  the order of 200
days or less. 

\section{Statistical self-similarity}

A set of time series $\delta p(t,\tau)$ is obtained for each value
of nonoverlapping time lag $\tau$.  The return of the stochastic
variable $\delta p(t,\tau)$ is said to be self-similar with
parameter $\alpha$ $(\alpha \geq)$, if for any $\lambda$
\begin{equation}
 \delta p(\tau)  \stackrel{EL}{=} \lambda ^{\alpha} \delta
p(\lambda\tau).\ \label{eq}
\end{equation}
The relation (\ref{eq}) is interpreted as an \textit{equality in
law} (EL), that is the two sides of the equation have the same statistical
properties. For the associated cumulative probability distribution, it
follows that
\begin{equation}
\wp(\delta p(\tau)\leq \rho)=\wp(\lambda ^{\alpha}\delta p(\lambda
\tau)\leq \rho), \ \label{cum}
\end{equation}
for any real $\rho$. This implies for the probability density $P$
\begin{equation}
P[\delta p(\tau)]=\lambda ^{-\alpha}P_{s}[\lambda^{-\alpha}\delta
p_{s}], \ \label{prob}
\end{equation}
introducing the master PDF $P_{s}$ with $\delta p_{s}= \delta
p(\lambda \tau)$. According to Eq. (\ref{prob}), there is a family
of PDFs that can be collapsed to a single curve $P_{s}$, if $\alpha$
is independent of $\tau$. This is known as monoscaling  in contrast
to multifractal scaling observed, e.g., for oil price returns at
macro-scales.

\begin{figure}
\includegraphics[width=0.5\textwidth]{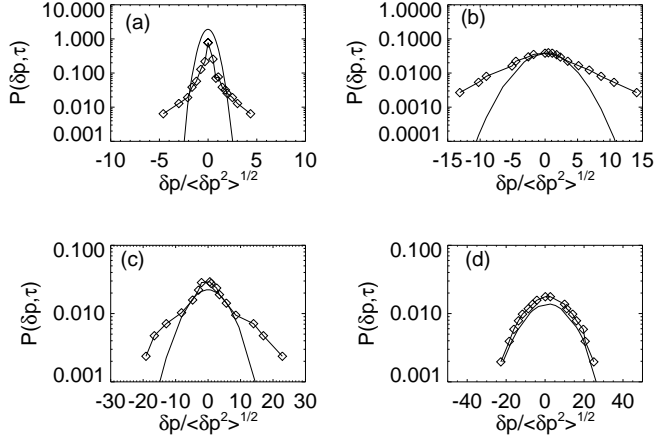}
\caption{\label{4_pdf} The PDFs of the normalized oil price
fluctuations for four different scales, $(a)$: $\tau=1$, $(b)$:
$\tau=20$, $(c)$: $\tau=60$, and $(d)$: $\tau=200$. For comparison,
the Gaussian distribution with the same variance is depicted(solid
line) }
\end{figure}

\begin{figure}
\includegraphics[width=0.5\textwidth]{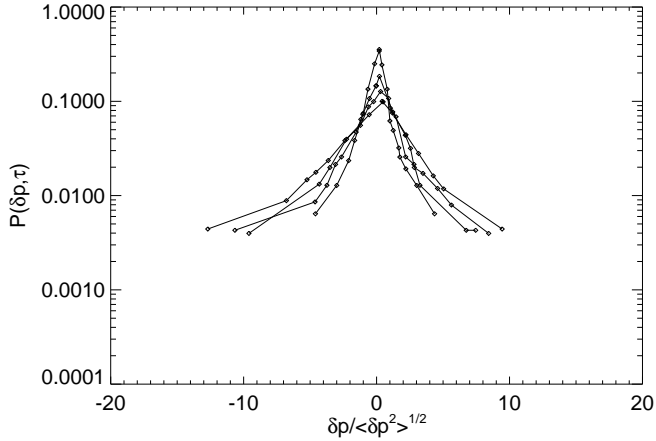}
\caption{\label{micro} The PDFs of the oil price fluctuations on
micro-scales.  We can see that the shape  of the PDFs do not change
fundamentally as a sequent   of the monofractality.}
\end{figure}

\begin{figure}
\includegraphics[width=0.5\textwidth]{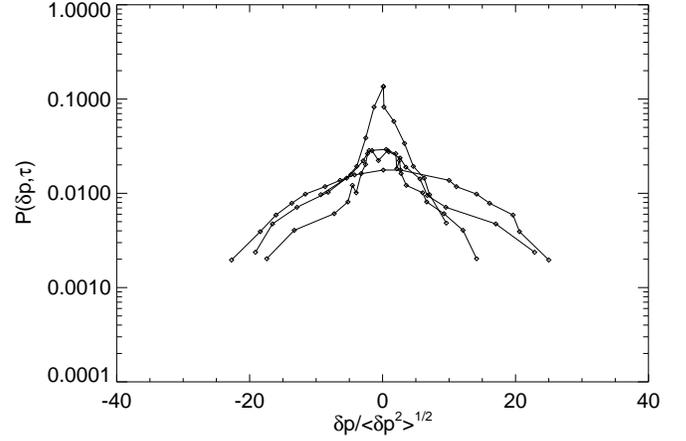}
\caption{\label{macro}  The PDFs of the oil price fluctuations on
macro  scales. The PDFs  become Gaussian as  the time scales is
increased. This is  coincide with multifractality  feature.}
\end{figure}

To characterize quantitatively the observed stochastic process, we
measure the PDF $P(\delta p)$ of the price fluctuations for $\tau$
ranging from 1 to 200 days. The number of data in each set decreases
from the maximum value of 5695 $(\tau=1)$ to the minimum value of
5495 $(\tau=200)$. In Fig.(\ref{4_pdf}) we show the four selected
PDFs (normalized with the variance $<\delta p(\tau) ^{2}>^{1/2}$)
for $\tau=1$, $\tau=20$, $\tau=60$, and $\tau=200$ from the top (
left side) to the bottom (right side) respectively. The
distributions lose their leptokurtic shape, as $\tau$ increases. Due
to the lack in  correlation among distant fluctuations, the
associated distributions become approximately Gaussian at
macro-scale. The scaling behavior of the distribution at coarser
time scales has two different regimes. At micro-scales (typically
shorter than 10 days), correlations between successive price changes
are dominant. This may be  due to several reasons, such as oil pipe
line damage, weather changes, or local variations in the internal (US) oil availability. 
Interestingly, the PDFs at the micro-scales have the same
leptokurtic shape, exhibit monoscaling, and do not change
fundamentally, and resemble closely L\'evy distributions, see Fig.
(\ref{micro}). On the other hand, at macro-scale (typically larger
than one month) permanent crisis in the Middle East and north
African govern the price drift and corresponding to a Gaussian
regime. This is coincide with a multifractal feature, as we can see
in Fig. (\ref{macro}).  However the micro-and macro time scales
regimes
  can be led to a linear and nonlinear scaling-dependence respectively.


First, let us   consider the scaling as defined by the structure
functions. The generalized structure function of order $n$ are
simply defined as
\begin{equation}
S^{n}(\tau;\pm \infty)=\langle |\delta
p|^{n}\rangle=\int_{-\infty}^{+\infty}|\delta p|^{n}P(\delta p,
\tau) d(\delta p) .\ \label{struc}
\end{equation}
The analysis which follows is also valid for the moments; however,
structure functions are typically calculated for a data series. The
arguments do not apply to structure functions of odd order, which
not only may have negative coefficients, but could in fact even change
the sign of the scaling range. The proof will, however, remain valid for
odd orders if the structure functions are defined with the
absolute value of the returns. Using the relation (\ref{prob}) we obtain
\begin{equation}
S^{n}(\tau;\pm \infty)=\lambda^{\zeta_{n}}S^{n}_{s}(\delta p_{s};\pm
\infty) , \ \label{struc1}
\end{equation}
where the linear function $\zeta_{n} = \alpha n$  refer to  the
statistical self-similarity,  monoscaling case.
  On the
contrary, in some cases, one may observe multifractality scaling, in
the sense that a nonlinear dependence is observed on $n$ where
$\zeta_{n}=n\alpha(n)$ is a convex function of $n$ and
$\zeta_{n+1}>\zeta_{n} \forall n$. This deviation from strict
self-similarity over all time scale $\tau$, also termed multifractal
scaling, is caused by the intermittent
micro-scale  structure of turbulence.

\begin{figure}
\includegraphics[width=0.5\textwidth]{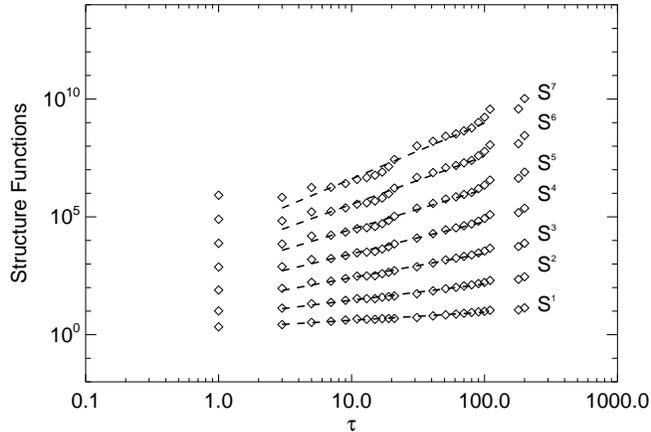}
\caption{\label{sf} The structure functions are depicted, as computed from Eq.
(\ref{struc}). In order to obtain the scaling exponents, we take the
logarithmic slope of linear least-square fits (solid line).}
\end{figure}

\begin{figure}
\includegraphics[width=0.5\textwidth]{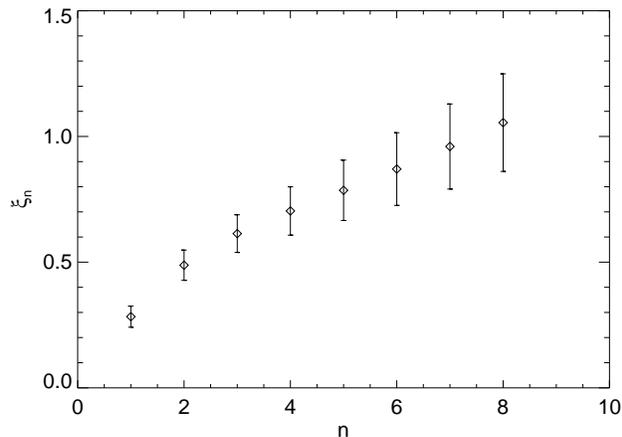}
\caption{\label{scaling} The scaling exponents $\zeta_{n}$ of the
structure functions are depicted versus the corresponding order $n$. The nonlinear
behavior is obvious in the figure.}
\end{figure}

To test if the above-mentioned interesting observations in  oil
price are a phenomenon related to inherent properties of stochastic
processes, structure functions $S^{n}(\tau)$ for different $\tau$, given by
Eq. (\ref{struc}),  are computed. A difficulty that can arise in the
experimental determination of the $\zeta_{n}$ is that for a finite
length times series, the integral Eq. (\ref{struc}) is not sampled
over the range $(-\infty;+\infty)$, rather  the outlying measured values of
$y$ determine the limit, $[-y ; +y]$. Fig. (\ref{sf}) shows some
selected $S^{n}(\tau)$, according to Eq. (\ref{struc}). The slope of
the curve gives scaling exponent $\zeta_n$ which can be obtained by
fitting a straight line to a log-log plot in the interval  $l\in
[1,100]$. Because of increasing of the statistical errors at the
higher orders, such a fitting becomes rather arbitrary. In Fig.
(\ref{scaling}) we report the scaling exponents extracted from the
structure functions. The behavior of $\zeta_{n}$ against $n$ shows
that scaling exponents have nonlinear behavior at all scales , say,
they are different from the usual  linear $\alpha n$ law. \\

\begin{figure}
\includegraphics[width=0.5\textwidth]{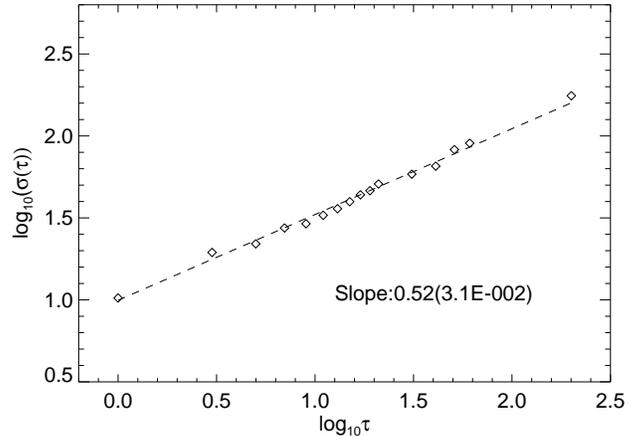}
\caption{\label{standard} Standard deviation of oil price returns
within the desired range. }
\end{figure}

To apply the rescaling procedure given by Eq. (\ref{prob}) the
exponent $\alpha$ is extracted from the underlying data by a
independent technique, \cite{kian,mom}. The standard deviation which
is defined by the root of the second-order structure function,
$\sigma(\tau)=[S^{2}(\tau)]^{1/2}$ and  has the minimum of
statistical error, exhibits power-law behavior with respect to the
increment distance, $\sigma(\tau)\sim \tau^{\alpha}$ as depicted in
Fig. (\ref{standard}).

\begin{figure}
\includegraphics[width=0.5\textwidth]{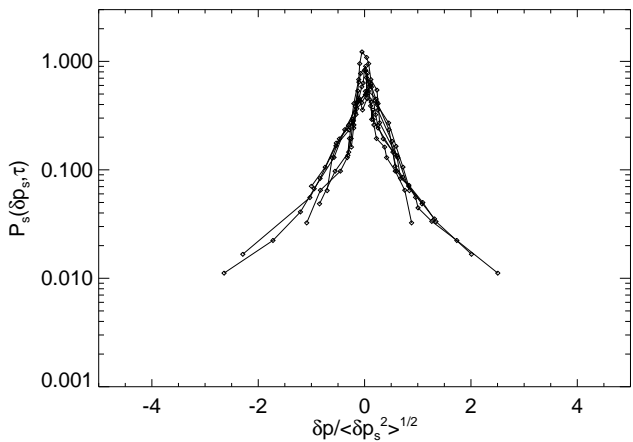}
\caption{\label{rescaled_macro} Rescaled PDFs of oil price
fluctuations in a macro-scale. This shows that the collapse of
the PDF's to a single curve is broken over the scales.}
\end{figure}

A linear least-square fit is carried out to obtain $\alpha$. The
characteristic exponent deduced in this way is $\alpha=0.52 \pm
0.031$ which is in good agreement with the value of $H$ obtained in
Section. (\ref{MF}). Fig. (\ref{rescaled_levy}) and Fig.
(\ref{rescaled_macro})
 show the rescaled sets of PDFs on the micro- and
macro-scales respectively. Evidently the PDFs at micro-scales  are
self-similar and collapse with weak scattering on the master PDF,
$P_{s}$, when using the characteristic exponents given above. The
corresponding increment distances $\tau$ are all shorter  than 10
days. We may model this PDFs by a L\'evy distribution, which thus turns out to be a
successful  fit to the distribution of oil price fluctuations. On the
other hand, at macro-scale the PDFs do not show a self-similar
behavior and rather constitute a multicascade process. This occurs at
scales larger than the one month. Because of the resulting
multifractal scaling of the PDFs it is evident that they can not
collapse onto a single curve, see Fig. (\ref{rescaled_macro}).

\section{L\'evy distribution model}

 L\'evy-stable laws are a rich
class of probability distributions that comprise   fat tails and
have many intriguing mathematical properties. They have been
proposed as models for many types of physical and economic systems.
There are several reasons for using L\'evy-stable laws to describe
complex systems. First of all, in some cases there are solid
theoretical reasons for expecting a non-Gaussian L\'evy stable
model, which can be a good fitting to experimental and numerical
results. The second reason is the Generalized Central Limit Theorem
which states that the only possible non-trivial limit of normalized
sums of independent identically distributed terms is L\'evy-stable.
(Recall that the classical Central Limit Theorem states that the
limit of normalized sums of independent identically distributed
terms with finite variance is Gaussian.) The third argument for
modeling with L\'evy-stable distributions is empirical; many large
data sets exhibit fat tails (or heavy tails);
 for a review see \cite{jani,silv}. Such data sets were  described by a Casting
 model \textbf{based on the log–normal ansatz (in terms of the variance of the Gaussian distribution)}. To confirm that
 model it is convenient to summarize the basic features of the L\'evy stable
 distribution.

 \begin{figure}
\includegraphics[width=0.5\textwidth]{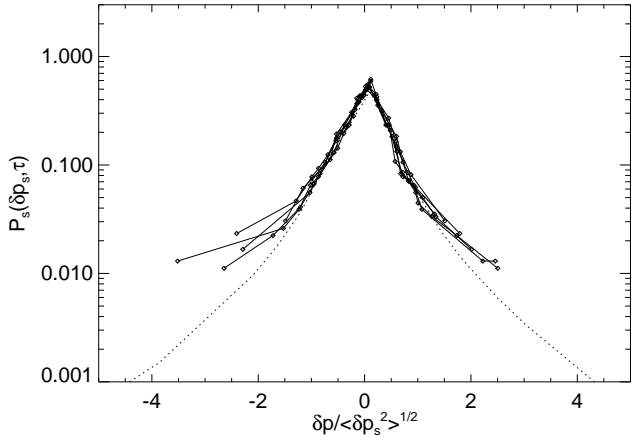}
\caption{\label{rescaled_levy}  Rescaled PDFs of oil price
fluctuations in micro-scales . The L\'evy law is represented by
the dashed line.}
\end{figure}

A L\'evy process is a time-dependent or position-dependent process
that at an infinitesimal interval has the L\'evy distribution of the
process variable. The characteristic function of the L\'evy process
is
\begin{equation}
K_{\mu}(q,s)=\exp(-cs\mid q \mid^{\mu})\, ,\label{levy111}
\end{equation}
where $s$ can be a characteristic time or space  scale. If $\mu=2$  the L\'evy
collapses to the Gaussian distribution. If $\mu=1$ the L\'evy
becomes a Cauchy distribution. The original L\'evy process is given
by its inverse Fourier transform, i.e.
\begin{equation}
P_{\mu}(x,s)=\int dq e^{iqx-cs\mid q \mid ^{\mu}}\, ,\label{levy123}
\end{equation}
and  the symmetric L\'evy distribution becomes
\begin{equation}
L_{\mu}(\delta x_{\Delta s})\equiv\frac{1}{\pi}\int_{0}^{\infty}
\exp(-\gamma \Delta s q^{\mu})\cos(q\delta x_{\Delta s})dq \,
,\label{levy1222}
\end{equation}
where the increment is $\delta x=x_{s}-x_{s-\Delta s}$; here, $0 < \mu < 2$, and
$\gamma>0$ is a scale factor. The maximum event probability leads to
\begin{equation}
P({0})=L_{\mu}(0)=\frac{\Gamma(1/\mu)}{\pi \mu(\gamma \Delta
s)^{1/\mu}}\, .\label{levy101}
\end{equation}
The exponent $\mu$ of the best fits is constant at  the  micro-scales range and amounts approximately to  $\mu \sim 1.92$ which is
$\mu=1/\alpha$. Natural phenomena also
investigated where similar findings have been reported include, for example, financial systems e.g. the Tehran price stock market, where $\mu \sim 1.36$ \cite{noroz}, and also physical systems such
as the solar wind, where $\mu \sim 3.3$ \cite{hnat}. From Fig.
(\ref{rescaled_levy}) we conclude that a central section of L\'evy
distributions describe very well the dynamics of the PDFs of oil price
fluctuations at micro-scales. One can see that the rescaled
PDFs are definitely non-Gaussian.

\section{Summary}

In this paper, we have presented a statistical analysis of  oil price in the United States for the
period of 1986 to 2008. We have applied a generic MF-DFA method to
extract scaling exponent of the fluctuation functions, in particular, relying on the
 second order $F_{2}(s)$ which was used in the rescaling procedure.
 However, the simple scaling properties that we have found via an
analysis of the PDFs, allow us to detect mono(multi) fractality
feature over all time scales. The presence of intermittency in  oil
price fluctuations is manifested by the leptokurtic nature of the
PDFs which show increased probability of large fluctuations compared
to that of the Gaussian distribution. Fluctuations on the macro
temporal scales, $\tau > 10 $ day, converge  toward  a Gaussian
distribution and are an uncorrelated signal. The reason may be due
to unstable conditions in the Middle East or OPEC's decisions oil production.
The critical  macro scale  which was  obtained is
different for some financial and physical systems; see for example in Refs.
\cite{noroz,ken}. We have also obtained   a good collapse onto a
single curve for  $\tau< 10 $, according to the rescaling procedure
(\ref{prob}). The proximity of the PDFs to L\'evy distributions is
made plausible by a simple model mimicking nonlinear spectral
transfer.

\bibliography{apssamp}

\end{document}